\documentclass[reprint,aps, prl, amsmath, amssymb]{revtex4-1}
\usepackage[utf8]{inputenc}
\usepackage{amsmath}
\usepackage{bm}
\usepackage{dcolumn}
\usepackage{natbib}
\usepackage{graphicx}
\usepackage[colorlinks, linkcolor= blue, citecolor = blue, urlcolor=blue]{hyperref}

\def\be{\begin{equation}}
\def\ee{\end{equation}}
\def \bea{\begin{eqnarray}}
\def \eea{\end{eqnarray}}
\def \nn{\nonumber}

\begin{document}

\title{Intrinsic Hall conductivities induced by the orbital magnetic moment}
\author{Kamal Das}
\email{kamaldas@iitk.ac.in}
\affiliation{Department of Physics, Indian Institute of Technology Kanpur, Kanpur 208016}
\author{Amit Agarwal}
\email{amitag@iitk.ac.in}
\affiliation{Department of Physics, Indian Institute of Technology Kanpur, Kanpur 208016}

\begin{abstract}
The intrinsic anomalous Hall effect is one of the most exciting manifestations of the geometric properties of the electronic wave-function.  Here, we predict that the electronic wave-function's geometric nature also gives rise to a purely quantum mechanical, {\it intrinsic} (scattering time independent) 
component of the Hall conductivities in the presence of a magnetic field. We show that the orbital magnetic moment and the anomalous Hall velocity combine to generate a scattering time independent contribution to the Hall effect, in addition to the Lorentz force induced scattering time dependent `classical' Hall effect. This dissipation-less Hall effect also manifests in the thermo-electric and thermal conductivities and is dominant near the band edges and band-crossings. It gives rise to negative magneto-resistance and it also leads to an underestimation in the charge carrier density measured in Hall experiments. 
\end{abstract}

\maketitle

The Hall effect is the generation of a voltage transverse to the applied current, and it can arise from several different mechanisms~\cite{Hall1879, Nagaosa10, Xiao06, Sodemann15, Papaj19, Mandal20}. The most prominent among them is the classical Hall effect, which arises in the presence of a magnetic field from the Lorentz force on the Bloch electrons~\cite{Hall1879, Hall1880}. 
Another fascinating Hall effect is the anomalous Hall effect~\cite{Karplus54, Sinitsyn08, Nagaosa10, Zhang16} that occurs in the absence of a magnetic field in time reversal symmetry (TRS) broken systems. Its intrinsic part arises from the Berry curvature (BC), a geometrical property of the electronic wave-function~\cite{Culcer03, Haldane04}, and it is used as a tool to probe the topological properties of material. This scattering-time independent anomalous Hall effect is quite counter-intuitive since transport signatures are mostly dissipative (scattering time dependent). In this letter, we explore another novel intrinsic 
Hall effect in the presence of a magnetic field, generated by the combination of the BC induced anomalous velocity and the orbital magnetic moment (OMM)~\cite{Chang96,Cooper97,Chang08, Thonhauser11, Ma15, Zhong16, Morimoto16a, Sekine18, Zhou19, Knoll19, Das19b, Gao19, Aryasetiawan19, Xiao20}. 

\begin{figure}
	\includegraphics[width=.97\linewidth]{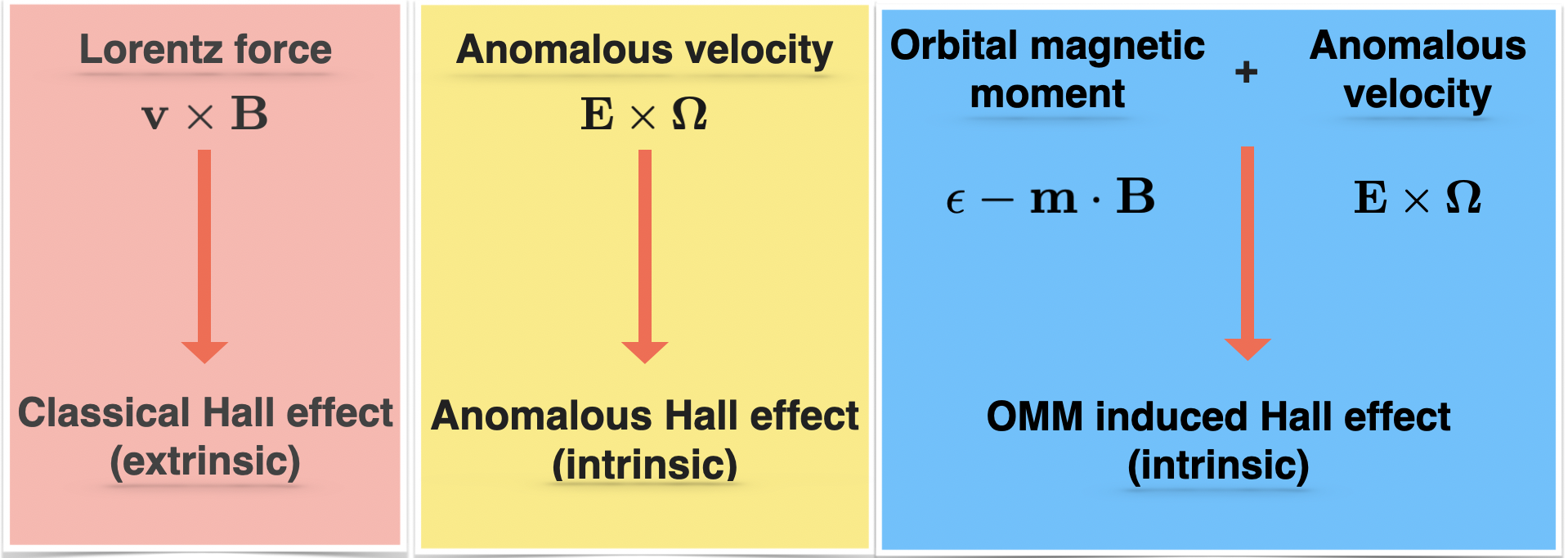} 
	\caption{The classical Hall effect arises from the Lorentz force which depends on the band velocity. 
	The anomalous Hall effect originates from the BC dependent anomalous velocity. The OMM-induced Hall effect arises from the combination of band energy modification by the OMM and the BC induced anomalous velocity. 
	\label{fig1}}
\end{figure}

A simple way to understand the OMM-induced intrinsic Hall effect [see Fig.~\ref{fig1}] is to start with the well known expression for the intrinsic anomalous Hall effect, 
\be  
\label{eq1}
\sigma_{ij}^{\rm A} =- \dfrac{e^2}{\hbar} \epsilon_{ijl} \int [d{\bf k}] \Omega_l f~.
\ee
Here, `$-e$' is the electronic charge, $\epsilon_{ijl}$ is the Levi-Civita tensor, $\Omega_l$ denotes the $l$ component of the BC and $[d{\bf k}] = d{\bf k}/(2\pi)^{d_0}$ in $d_0$ dimensions. We have defined $f  = f[\beta (\epsilon - \mu)]$ with $f(x) = 1/(1+ e^x)$ as the Fermi-Dirac distribution function for the concerned band with dispersion $\epsilon$ and inverse temperature $\beta = 1/(k_B T)$. In materials with intrinsic TRS, the BC is an odd function of the crystal momentum and thus the anomalous Hall effect vanishes when summed over the Brillouin zone. However, the OMM (${\bf m}$) modifies the electronic band dispersion in the presence of magnetic field ${\bf B}$, through a Zeeman like coupling $\epsilon\to \tilde{\epsilon}= \epsilon - {\bf m}\cdot {\bf B}$. Thus, to lowest order in the magnetic field, Eq.~\eqref{eq1} reduces to 
\be  \label{Hall_cond}
\sigma_{ij}^{\rm O} = \dfrac{e^2}{\hbar} \epsilon_{ijl} \int [d{\bf k}] \Omega_l \epsilon_{m} ~ f^\prime~,
\ee
with $\epsilon_{m}={\bf m} \cdot {\bf B}$ and $f^\prime = \partial_{\epsilon} f$. 
This dissipation-less Hall effect also manifests in other thermo-electric and thermal transport phenomena such as the Nernst effect, the Ettinghausen effect and the  thermal Hall effect. 
These contributions are purely quantum mechanical in nature and have no classical analogue. 
In non-centrosymmetric materials with TRS, the OMM-induced Hall conductivity, specified by Eq.~\eqref{Hall_cond}, is the only intrinsic Hall response. 

In contrast to the classical Hall conductivities, the OM-induced Hall conductivities dominate near the band edges or band crossings, which are also the `hotspot' of the BC and the OMM. Using gapped graphene and Weyl metals as specific examples for 2D and 3D systems, we demonstrate that these OM-induced Hall components have a significant impact on the magneto-resistance (MR) measurements, giving rise to a negative MR~\cite{Zhou19}. Most importantly, we show that the additional 
OMM-induced Hall conductivity contribution leads to a significant under-estimation in the charge carrier density measured in conventional Hall resistivity experiments. 

The phenomenological transport equations for the Hall response in charge current (${\bf j}$) and thermal current (${\bf J}$) are given by
\be \label{Hall_conductivities}
\begin{pmatrix}
j_x \\ J_x
\end{pmatrix} =
\begin{pmatrix}
\sigma_{xy} & \alpha_{xy} \\
\bar{\alpha}_{xy} & \bar{\kappa}_{xy} 
\end{pmatrix}
\begin{pmatrix}
E_y \\ -\nabla_y T
\end{pmatrix}~.
\ee
Here, ${\bf E}$ denotes the electric field and ${\bf \nabla}T$ is the temperature gradient. 
The conductivities $\sigma_{xy}$ and $\alpha_{xy}$ ($\bar \alpha_{xy}$ and $\bar \kappa_{xy}$) capture the charge (heat) flow perpendicular to the applied electrical and thermal bias. We will refer to all these collectively as Hall conductivities. Other experimentally relevant transport quantities such as the Nernst, Seebeck, and thermal conductivity can be expressed in terms of these.

In presence of the BC, the magneto-transport in the semiclassical regime is described by three ingredients: 
i) the equations of motion of the carriers, ii) the modified momentum  space volume for the  Brillouin zone (BZ) integration and iii) the non-equilibrium distribution function. In the `normal' Hall configuration~\cite{Hall1879} (${\bf E} \perp {\bf B}$), the equation of motion describing the position ({\bf r}) and momentum ($\hbar {\bf k}$) dynamics of the carrier wave-packet is given by~\cite{Sundaram99, Xiao10}
%
%
\bea  \label{band_velocity}
\hbar {\bf \dot k} &=& D \left[-e {\bf E} -  e(\tilde{\bf v} \times {\bf B})  \right]~,
\\
\hbar {\bf \dot r} &=& D\left[\hbar \tilde{\bf v} + e{\bf E} \times  {\bf \Omega}+ e(\tilde{\bf v} \cdot {\bf \Omega}){\bf B}\right]~.
\eea
%
Here, $D$ is the phase-space factor~\cite{Xiao05} defined as $1/D = \left[1 + \frac{e }{\hbar}({\bf B} \cdot {\bf \Omega}) \right]$. The OMM modified band velocity is given by $\hbar \tilde {\bf v} =\nabla_{\bf k} \tilde{\epsilon}$. Consequently, the carrier velocity can be expressed as $\tilde {\bf v}={\bf v} - {\bf v}_m$ where the band velocity is $ \hbar {\bf v} =  \nabla_{\bf k} \epsilon$ and the OMM-induced velocity $\hbar {\bf v}_{m} = \nabla_{\bf k} \epsilon_{m}$. In presence of BC and magnetic field, the phase-space volume is modified by $d{\bf k} \to [D]^{-1} d{\bf k}$. 
For calculating the non-equilibrium distribution function, $g_{{\bf r},{\bf k}}$, we use the semiclassical Boltzmann equation given by ~\cite{Ashcroft76}
\begin{equation}\label{boltzmann}
\dfrac{\partial g_{{\bf r},{\bf k}}}{\partial t} + \dot{\bf k} \cdot \nabla_{\bf k} g_{{\bf r},{\bf k}}+ \dot{\bf r} \cdot \nabla_{\bf r} g_{{\bf r},{\bf k}} =- \frac{g_{{\bf r},{\bf k}} - \tilde f}{\tau}~.
\end{equation}
Here, $ \tilde f = f[\beta(\tilde \epsilon - \mu)]$ 
and $\tau$ is the scattering time. In the steady state regime with bias, we have $g_{{\bf r},{\bf k}} \approx \tilde f + \delta g_{{\bf r},{\bf k}}$, where $\delta g_{{\bf r},{\bf k}}$ is the non-equilibrium part.  To the lowest order in external biases and magnetic field, in the normal Hall configuration [$({\bf E} ~\mbox{and}~ \nabla_{\bf r} T) \cdot {\bf B}=0$], $\delta g_{{\bf r},{\bf k}}$ can be calculated to be 
\bea \nn
\delta g&=&  \tau \Big[  \tau \frac{e}{\hbar} \{ ( {\bf v} \times {\bf B})\cdot \nabla_{\bf k}\} f^\prime {\bf v}  - \Omega_{\rm  B}  f^\prime  {\bf v}  -  f^\prime {\bf v}_m - \epsilon_m f^{\prime \prime} {\bf v}  \Big] 
\\ 
&&  \cdot \left [e{\bf E} +\beta \left(\epsilon - \mu\right)k_B \nabla_{\bf r}T\right] - \tau f^\prime \beta \epsilon_m {\bf v}\cdot k_B \nabla_{\bf r}T.~~ ~~
\eea
Here, we have defined, $\Omega_{\rm B} = e/\hbar \bf{\Omega} \cdot \bf{B}$ and $f^{\prime \prime} = \partial^2_{\epsilon} f$. Using these in conjugation  with the proper definition of the physically observable current, which excludes the 
OMM-induced circulating currents (see Supplementary material (SM) 
\footnote{\href{https://www.dropbox.com/s/3dlnobd9nm195t6/SM_OMM_HE.pdf?dl=0}{
Supplementary material detailing semiclassical calculation for i) intrinsic off-diagonal, ii) extrinsic off-diagonal, and  iii) extrinsic diagonal components of the conductivity, iv) Hall coefficients, v) magneto-resistance, and vi) Weyl metals.}} 
for details), 
all the conductivities can be calculated. 

If the magnetic field is applied along the $z$-direction then the conductivities defined in Eq.~\eqref{Hall_conductivities} represent the `normal' Hall conductivities~\cite{Hall1879}. We find that in an intrinsically TRS preserving system there are two distinct contributions of these. The well known Lorentz force contribution is $\propto \tau^2$, and are given by ~\cite{Hurd72, Ziman72, Auerbach18} 
\bea \label{lrntz}
\begin{pmatrix}
\sigma_{xy}^{\rm L} & \alpha_{xy}^{\rm L}  \\
\bar{\alpha}_{xy}^{\rm L}  & \bar{\kappa}_{xy} ^{\rm L} 
\end{pmatrix} 
& = &
- \dfrac{e^3 \tau^2}{\hbar}  \int [d{\bf k}] 
\begin{pmatrix}
1 & - \frac{\epsilon -\mu}{eT}\\
- \frac{\epsilon -\mu}{e}  & \frac{\left(\epsilon -\mu\right)^2}{e^2 T} 
\end{pmatrix}
 f^\prime 
\nn \\ 
 &\times & [v_x ({\bf v \times {\bf B}})\cdot {\nabla}_{\bf k} v_y ]~.
\eea
In addition to the Hall conductivity, the Lorentz force also modifies the longitudinal conductivity. This modification can be calculated to be $\delta \sigma_{xx}^{\rm L} = -(e^4 \tau^3/\hbar^2) \int [d{\bf k}] v_x 
[\{({\bf v}\times {\bf B})\cdot \nabla_{\bf k}\}^2 v_x] f^\prime$.
%

More interestingly, the OMM combines with the anomalous velocity (see Fig.~\ref{fig1}) to give rise to an additional {\it intrinsic}  component of the Hall conductivities. We calculate these OMM-induced conductivities to be  (see SM\cite{Note1} for details)
\be \label{sigma_orb}
\begin{pmatrix}
\sigma_{xy}^{\rm O} & \alpha_{xy}^{\rm O}  \\
\bar{\alpha}_{xy}^{\rm O}  & \bar{\kappa}_{xy} ^{\rm O} 
\end{pmatrix} 
= 
\dfrac{e^2}{\hbar}  \int [d{\bf k}] \epsilon_{ m} \Omega_z
\begin{pmatrix}
1 & - \frac{\epsilon -\mu}{eT}\\
- \frac{\epsilon -\mu}{e}  & \frac{\left(\epsilon -\mu\right)^2}{e^2 T}~
\end{pmatrix}
 f^\prime~.
\ee
The prediction of these {\it intrinsic} OMM-induced Hall conductivities is one of the main highlights of this paper. These dissipation-less components are of purely quantum mechanical origin and have no classical counterpart. In this letter, we will focus on non-magnetic systems which break inversion symmetry and intrinsically preserve the TRS. The case of magnetic systems which intrinsically break the TRS is discussed in detail in the SM~\cite{Note1}.

In addition to the Hall components, the BC and the OMM also give rise to modifications in the longitudinal conductivities. The longitudinal component of the charge conductivity is given by~\cite{Das19b}
\bea \label{delta_sigma_xx}\nn
\delta \sigma_{xx}^{\rm O} = - e^2 \tau \int [d{\bf k}] \Big[ v_x^2 \Omega_{\rm B}^2f^\prime  + v_x  \left( \epsilon_m f^{\prime \prime} + 2 \Omega_{\rm B}f^{\prime}\right) v_{mx}
\\
 + v_x^2 (\epsilon_m^2 f^{\prime \prime \prime}/2 + \Omega_{\rm B}\epsilon_m f^{\prime \prime})+ v_{mx} ( f^\prime v_{mx} + \epsilon_m f^{\prime \prime} v_x )\Big].~~~~
\eea
The details of the calculation and the longitudinal contributions in the thermo-electric and thermal conductivities are presented in the SM~\cite{Note1}.

We find that similar to the classical Hall conductivities, the OMM-induced Hall conductivities also satisfy the i) Onsager's reciprocity relation, ii) the Wiedemann-Franz law, and iii) the Mott relation. This ensures that the other three conductivities can be deduced from the Hall conductivity~\cite{Xiao06,Dong20}. More explicitly, if the Hall conductivity is known, then for $\mu\gg k_B T$, the Wiedemann-Franz law implies $\bar \kappa_{xy}^{\rm O} =  \pi^2 k_B^2 T\sigma_{xy}^{\rm O}/(3e^2)$, the Mott relation ensures $ \alpha_{xy}^{\rm O} =- \pi^2 k_B^2 T\partial_\mu \sigma_{xy}^{\rm O}/(3e)$, and the Onsagar's relation leads to $\bar \alpha_{xy}^{\rm O} = T \alpha_{xy}^{\rm O}$. 

\begin{table}[t]
	\caption{The calculated carrier density, BC, OMM, and OM-induced Hall and longitudinal conductivities in gapped graphene and in WM for conduction band ($\mu>0$). We have written the charge conductivity in units of $\sigma^{\rm GG} = g_s g_v \frac{e^2}{\hbar}$ for gapped graphene and in units of $\sigma^{\rm WM} = g_w \frac{e^2}{\hbar}\frac{ \mu}{ \hbar v_F}$ for WM. The thermoelectric conductivity is written in units of $k_B\sigma^{\rm GG}/e$ for the former and in units of $k_B \sigma^{\rm WM}/e$ for the latter.}
	\centering  
	{
		\begin{tabular}{c  c  c}
			\hline \hline
			\rule{0pt}{3ex}
			   &  Gapped graphene & Weyl metal \\ [1ex]
			\hline \hline
			\rule{0pt}{3ex}
			$n$ ~~~&~~~ $g_s g_v \left( \mu^2 -\Delta_g^2\right)/(4\pi  \hbar^2 v_F^2)$~~~~ & ~~~~$g_w \mu^3/ (6 \pi^2 \hbar^3 v_F^3)$\\ [2ex] 
			${\rm BC}$ & $-\tau_z \frac{\hbar^2 v_F^2 \Delta_g}{2 \epsilon^3} {\hat z}$  
			&~$- \chi \frac{\bf k}{ 2 |{\bf k}|^3}$  \\ [2ex]
			${\rm OMM}$ & $- \tau_z \frac{e \hbar v_F^2 \Delta_g}{2 \epsilon^2}{\hat z}$ & $ - \chi e v_F \frac{\bf k}{ 2 |{\bf k}|^2}$\\[2ex]
			 \rule{0pt}{3ex}
			 $\gamma_B$ &$\hbar v_F^2\Delta_ge B/(2\mu^3)$ & $\hbar v_F^2 e B/(2\mu^2)$\\ [2ex] 
			 \hline \rule{0pt}{3ex}
			$\sigma_{xy}^{\rm O}$ & $- \frac{1}{4\pi} \frac{\Delta_g}{\mu} \gamma_{\rm B}$ & $-  \frac{1}{12 \pi^2} \gamma_{\rm B}$~\\[2ex]
			$\delta\sigma_{xx}^{\rm O}$  & $-  \frac{ \mu \tau}{4 \pi \hbar} \left[5 - 2\Delta_g^2/\mu^2 \right] \gamma_{\rm B}^2 $~~~~~ & 
			$- \frac{\mu \tau}{ 15 \pi^2\hbar}\gamma_{\rm B}^2 $\\[2ex] 
			\hline
			\rule{0pt}{3ex}
			$\alpha_{xy}^{\rm O}$ & $ -\frac{\pi}{3\beta \mu}  \frac{\Delta_g}{\mu} \gamma_{\rm B}$& $-\frac{1}{36 \beta \mu }\gamma_{\rm B}$ \\ [2ex] 
			$\delta \alpha_{xx}^{\rm O}$ &  $- \frac{\pi \tau}{12\beta \hbar} \left[25 - 14\Delta_g^2/\mu^2 \right]\gamma_{\rm B}^2  $& $- \frac{2 \tau}{45 \beta \hbar} \gamma_{\rm B}^2 $ \\ [2ex] 
			$\alpha_{xy}^{\rm L}$ & $ \frac{\pi \tau}{6 \beta \hbar} \frac{\Delta_g^2}{ \mu^2 } (\omega_c \tau)$& $ \frac{\tau}{18 \beta \hbar} (\omega_c \tau)$ \\ [2ex] 
			$\delta \alpha_{xx}^{\rm L}$ &$\frac{\pi \tau}{12\beta \hbar }  (3\Delta_g^2/\mu^2 - 1 ) (\omega_c \tau)^2$ & $0$ \\ [2ex] 
			$\alpha_{\rm D}$ & $-\frac{\pi \tau}{12 \beta \hbar}(1+ \Delta_g^2/\mu^2)$ &$ -  \frac{ \tau}{ 9 \beta \hbar}$ \\ [2ex] 
			\hline \hline
	\end{tabular} }
	\label{table_1}
\end{table}

To explore the quantitative impact of the OMM-induced conductivities, we consider two experimentally interesting systems with broken inversion symmetry. In two dimension, we choose gapped graphene whose low energy physics is described by a massive Dirac Hamiltonian. Gapped graphene can be realized by a heterostructure of graphene on hexagonal boron nitride~\cite{Zhou07, Yankowitz12, Song15}. The valley resolved Hamiltonian of this system is given by~\cite{Xiao07} 
\be \label{ham_2}
{\mathcal H}=\hbar v_F(\tau_z k_x \sigma_x + k_y\sigma_y) + \Delta_g\sigma_z~.
\ee
Here, $\sigma_i$ are the Pauli spin matrices denoting the pseudo-spin, $\tau_z = \pm 1$ denotes the $K/K'$ valley degree of freedom and $2 \Delta_g$ is the bandgap at each valley induced by the breaking of the sub-lattice symmetry~\cite{Zhou07}. The bands for the two spins are degenerate  ($g_s=2$) and there are two valleys ($g_v=2$) in the Brillouin zone. In three dimension, we choose an inversion symmetry broken Weyl metal (WM). WMs can be realized in the inversion symmetry broken TaAs family~\cite{Huang15, Lv15, Armitage18}. A minimal model of such WM hosts two pairs of Weyl nodes of opposite chirality ($g_w=4$). The Hamiltonian for each of the Weyl node is~\cite{Das20a, Das20b}
\be 
\label{H_WSM}
{\mathcal H} = \chi \hbar v_F \left( \sigma_x k_x + \sigma_y k_y + \sigma_z k_z \right)~.
\ee
Here, $\chi$ denotes the chirality of the given Weyl node. 

We calculate the BC, OMM, and the OMM-induced electric and thermo-electric conductivities for both these systems [Eqs.~\eqref{ham_2} and \eqref{H_WSM}], and the results are tabulated in Table~\ref{table_1}. For the OMM-induced conductivities we have defined a dimensionless parameter $(\gamma_{\rm B}$) related to the OMM coupling energy scale at the Fermi surface, $\gamma_{\rm B} = |{\bf m}(\mu)| B/\mu$. The model specific expressions of $\gamma_B$ are also tabulated in Table~\ref{table_1}.  

The classical Hall conductivity for both these systems can be expressed as $\sigma_{xy}^{\rm L} =-\omega_c \tau \sigma_{\rm D}$, where the cyclotron frequency is $\omega_c=eB/m^*$ and the Drude conductivity is given by $\sigma_{\rm D} = n e^2 \tau /m^*$. The carrier density $n$ is specified in Table~\ref{table_1} as a function of $\mu$ and the inertial or cyclotron mass for both these isotropic systems is given by $m^* = \mu/v_F^2$~\cite{Sachdeva15}. 
\begin{figure}[t]
\includegraphics[width=.97\linewidth]{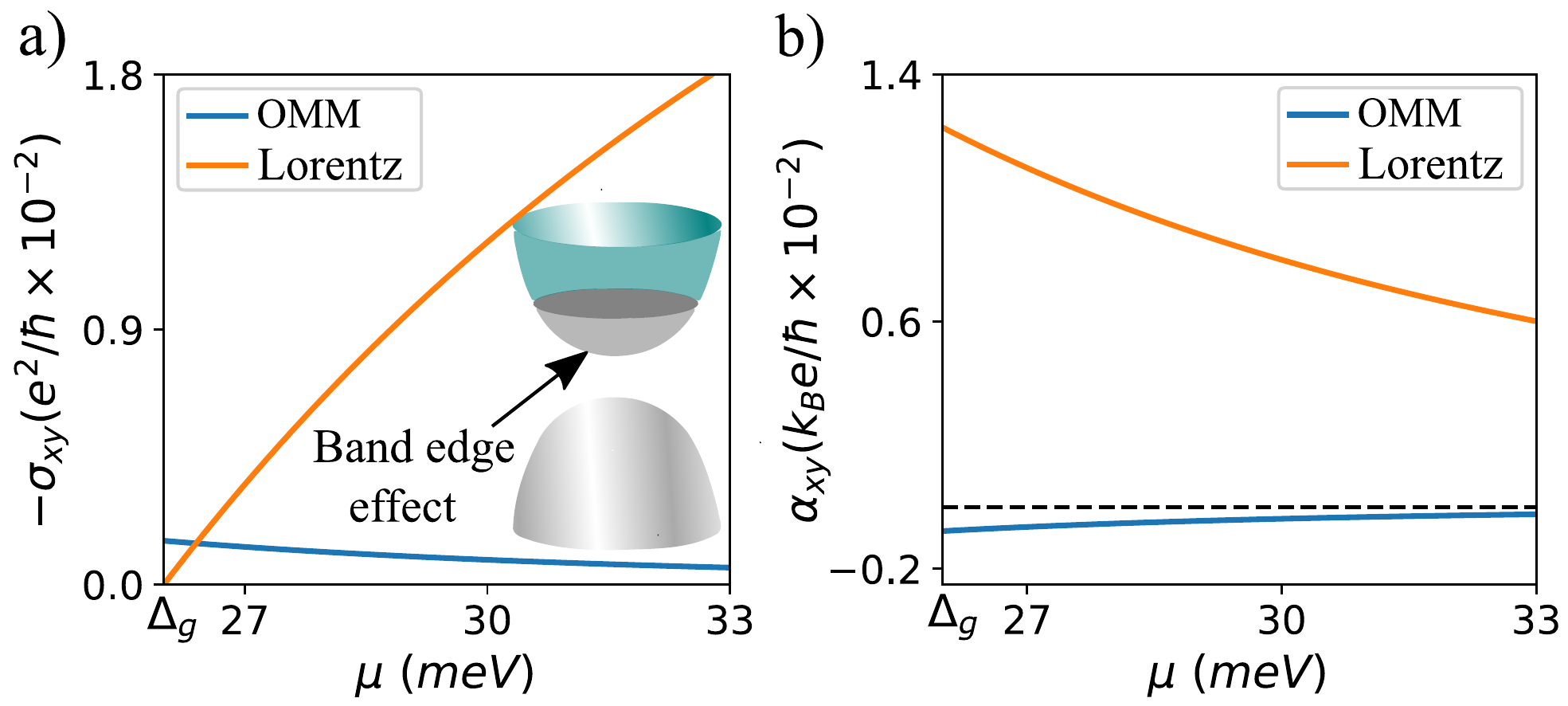} 
\caption{The OMM-induced Hall effect (blue line) and the Lorentz force induced classical Hall effect (orange line) for gapped graphene. (a) The variation of $\sigma_{xy}$ with $\mu$ in the conduction band. The OMM contribution is peaked near the band edge. (b) The variation of $\alpha_{xy}$ with $\mu$ at $15$ K. Both the components of $\sigma_{xy}$ have the same sign, while the OMM-induced $\alpha_{xy}^{\rm O}$ has the sign opposite to that of the Lorentz contribution. We have chosen the other parameters to be  
$\Delta_g = 26$ meV \cite{Zhou07, Yankowitz12},  $v_F = 10^5$ m/s, $B =1$ T and $\tau =0.1$ ps~\cite{Zhou19}.
\label{fig_2}}
\end{figure}
We find that chemical potential dependence of $\sigma_{xy}^{\rm O}$ is opposite to its classical counterpart. $\sigma_{xy} ^{\rm O}$ is largest near the band-edge or the band crossing point, the `hotspots' of the BC and the OMM, and it decreases as we move away from the band-edge to regions of larger carrier density. 
This is highlighted in Fig.~\ref{fig_2}(a). Depending on the value of the scattering timescale, the OMM contribution can be of the same order as the classical Hall contribution in the low carrier region near the band edge. In the thermoelectric Hall conductivity, the magnitude of both the Lorentz and the OMM contributions are peaked near the band-edge and they decrease as we move away from the band edge. We find that similar to the Lorentz contribution, $\sigma_{xy}^{\rm O}$ retains the sign of the charge carriers, changing from positive to negative as we move from the conduction band (electrons) to the valance band (holes). However, $\alpha_{xy}^{\rm O}$ retains the same sign for both the bands. 


These relatively unexplored OMM-induced Hall conductivities have a significant impact on i) the Hall resistivity (or coefficient) which is used to determine the carrier density, and ii) the transverse MR. We discuss both of these below.

\begin{figure}[t]
\includegraphics[width=.97\linewidth]{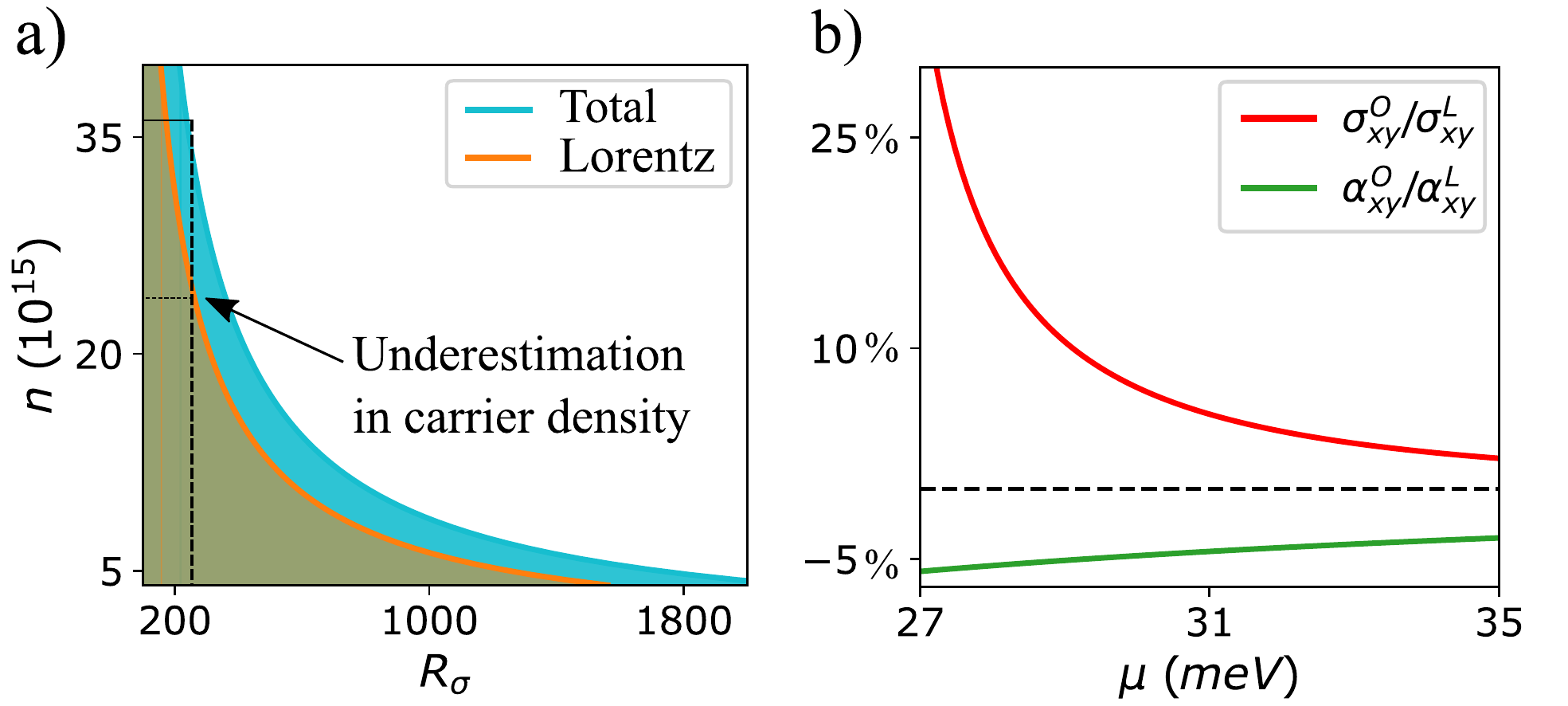} 
\caption{(a) The carrier density as a function of Hall coefficient for gapped graphene. The orange line shows the classical Hall coefficient and the cyan line represents the total Hall coefficient including the OMM correction. Since the $\sigma_{xy}^{\rm O}$ adds an extra contribution to the Hall coefficient, the measurement of the number density using the relation $R_{\sigma} = 1/(ne)$, leads to an underestimation of $n$.  (b) The ratio $\sigma_{xy}^{\rm O}/\sigma_{xy}^{\rm L}$ in red line and $\alpha_{xy}^{\rm O}/\alpha_{xy}^{\rm L}$ in green line. 
\label{fig_3}}
\end{figure}
%

Similar to the electron gas with parabolic dispersion, we find that the classical Hall resistivity [$\rho_{xy} = -\sigma_{xy}/({\sigma_{xx}^2 + \sigma_{xy}^2})$] in gapped graphene and in WM can be expressed as $\rho_{xy}^{\rm L} =  B/(ne)$ where $n$ is the carrier density~\cite{Ashcroft76}. Owing to this very simple carrier density dependence, the Hall resistivity measurement is routinely used to experimentally estimate the carrier density in terms of Hall coefficient ($R_\sigma$), 
\be \label{eqn_Rh}
n = \frac{1}{e} R_\sigma^{\rm L}~,~~~{\rm where}~~~ R_\sigma^{\rm L} \equiv \left. \frac{\partial\rho_{xy}^{\rm L}}{\partial B}\right|_{B\to 0}~.
\ee  
However, in presence of the OMM-induced Hall conductivity, the total Hall resistivity is given by $\rho_{xy}^{\rm total} = \rho_{xy}^{\rm L} \left[ 1 + \sigma_{xy}^{\rm O}/\sigma_{xy}^{\rm L}\right]~+ {\cal O}(B^3)$.
Thus, we have $R_\sigma^{\rm L} \to R_\sigma^{\rm total} $, where the total ($B$ independent) Hall coefficient is 
\be \label{R_sigma}
R_\sigma^{\rm total}= R_\sigma^{\rm L} \left( 1+ \dfrac{ \sigma_{xy}^{\rm O}}{\sigma_{xy}^{\rm L}}\right)~.
\ee
This additional contribution, $R_\sigma^{\rm O} = r_\sigma R_\sigma^{\rm L}  $, with $r_\sigma\equiv\sigma_{xy}^{\rm O}/\sigma_{xy}^{\rm L}$ leads to an underestimation of the carrier density that we extract from Hall measurement, particularly near the band edges as shown in Fig.~\ref{fig_3}. 

In gapped graphene, we find the OMM-induced {\it non-universal} correction to the the Hall coefficient to be 
\be 
r_\sigma^{\rm GG}= \frac{\sigma^{\rm O}_{xy}}{\sigma_{xy}^{\rm L}} = \dfrac{1}{2} \dfrac{\Delta_g^2}{(\mu^2 - \Delta_g^2)} \dfrac{\hbar^2}{\mu^2 \tau^2}~.
\ee
This correction explicitly depends on the ratio of scattering energy scale and the chemical potential and represents $35 \%$ underestimation of the carrier density, as shown in Fig.~\ref{fig_3}(a). 
A similar correction for the WM can be easily obtained to be $r_\sigma^{\rm WM} = {\hbar^2}/{(4\mu^2\tau^2)}$. For the typical choice of $\mu =25$ meV and $\tau=0.1$ ps, this turns out to be $2\%$ which leads to approximately $2\%$ underestimation of the carrier density, without explicitly accounting for the OMM correction 
\footnote{Similar OMM induced corrections can also be obtained for the Nernst number $R_\nu \equiv (\sigma_{xx} \bar \kappa_{xx})^{-1} \partial_B\alpha_{xy}|_{B\to 0}$ 
and the thermal Hall number~\cite{Auerbach19} $R_{\bar \kappa} \equiv (\bar \kappa_{xx})^{-2} \partial_B \bar \kappa_{xy}|_{B\to 0}$.
Both of these have the form $R_i =R_i^{\rm L} (1+ r_i )$ with $r_\nu  = \nu_{xy}^{\rm O}/\nu_{xy}^{\rm L}$ and $r_{\bar \kappa} = \bar \kappa_{xy}^{\rm O}/\bar \kappa_{xy}^{\rm L}$.
More importantly, the OMM contribution in these Hall coefficients are extrinsic in nature making them {\it non-universal}, in contrast to the Lorentz contribution which is intrinsic \cite{Auerbach18,Auerbach19}.
}.

\begin{figure}[t]
\includegraphics[width=.97\linewidth]{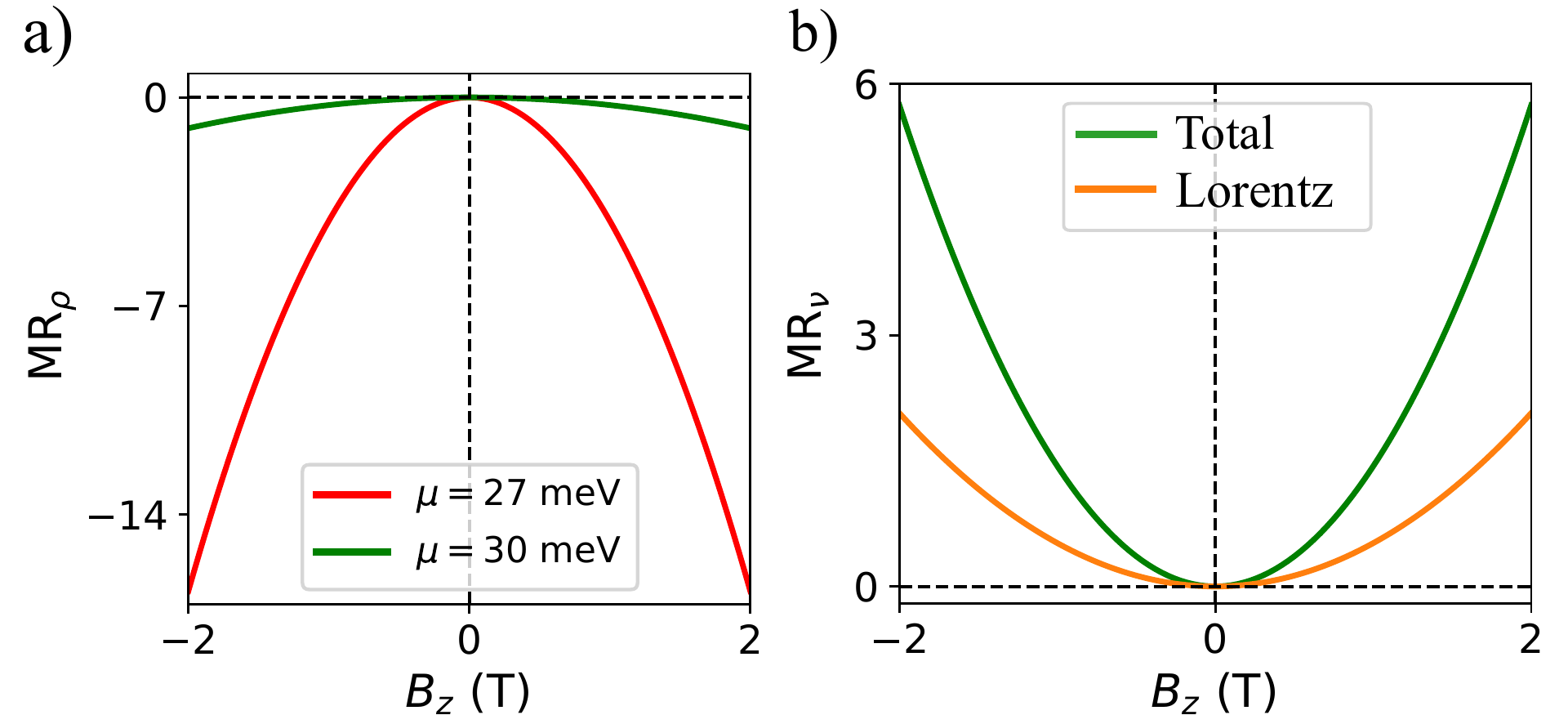} 
\caption{The magnetic field dependence of MR$_\rho$ and MR$_\nu$ in gapped graphene. (a) The OM-induced MR$_\rho$ for two different choice of chemical potential. (b) The total positive MR$_\nu$ (green), including all contributions, is significantly higher than its value obtained by considering only the Lorentz force contribution (orange curve) at $\mu=27$ meV. 
\label{fig_4}}
\end{figure}
%

The OMM-induced intrinsic Hall conductivities also manifest in the measured MR of resistivity (MR$_\rho$ for $\rho_{xx}$), MR of the Seebeck conductivity (MR$_\nu$ for $\nu_{xx} = \rho_{xx} \alpha_{xx} + \rho_{xy} \alpha_{xy}$) and in other transport measurements.

For the MR$_\rho \equiv \rho_{xx}(B)/\rho_{xx}(0) - 1$, the OMM-induced Hall conductivity can give rise to a negative MR \cite{Zhou19}. For an time-reversal symmetric system, to lowest order in the magnetic field, the MR can be calculated to be, 
\be \label{MR_rho}
{\rm MR}_{\rho} = - \left[ \frac{\delta\sigma_{xx}^{\rm O}}{\sigma^{\rm D}} + \frac{\sigma^{\rm O}_{xy}(\sigma^{\rm O}_{xy} + 2 \sigma^{\rm L}_{xy})}{(\sigma^{\rm D})^2}\right]~.
\ee
As expected, we find that the Lorentz force by itself cannot induce any MR. The {\it quadratic negative} MR$_\rho$ is maximum near the band edges where the Drude conductivity is minimal, and the OM-induced contributions are maximum. Here the second term is $\propto 1/\tau^2$ while the rest are independent of $\tau$.  We find that the OMM and BC induced longitudinal conductivity correction $\delta \sigma^{\rm O}_{xx}$, calculated from Eq.~\eqref{delta_sigma_xx} and tabulated in Table~\ref{table_1}, decrease the conductance for both gapped graphene and WM. Thus, the first term in Eq.~\eqref{MR_rho} leads to a positive MR$_\rho$ component. However, the OMM-induced Hall conductivity corrections are stronger and it makes the total MR$_\rho$ negative, as shown in Fig.~\ref{fig_4}(a). 

The MR in the Seebeck coefficient is defined as MR$_\nu \equiv \nu_{xx}(B)/\nu_{xx}(0) - 1$. For systems which intrinsically preserve TRS, to lowest order in magnetic field, we obtain 
\be \label{MR_nu}
{\rm MR}_{\nu} = {\rm MR}_\rho + \dfrac{\delta \alpha^{\rm L}_{xx} + \delta \alpha^{\rm O}_{xx} }{\alpha_{\rm D}} +\dfrac{\sigma_{xy}}{\sigma_{\rm D}} \dfrac{\alpha_{xy}}{\alpha_{\rm D}}.
\ee
Here $\sigma_{xy} = \sigma_{xy}^{\rm L} + \sigma_{xy}^{\rm O}$ and $\alpha_{xy} = \alpha_{xy}^{\rm L} + \alpha_{xy}^{\rm O}$. 
In contrast to its resistivity counterpart, the Lorentz force contribution alone ($\delta \alpha_{xx}^{\rm L}$, $\sigma_{xy}^{\rm L}$ and $\alpha_{xy}^{\rm L}$) causes a finite positive MR$_\nu$. We find that the OMM-induced corrections contribute to the MR$_{\nu}$, thus enhancing the strength of the total MR$_\nu$. This is shown explicitly for the case of gapped graphene in Fig.~\ref{fig_4}(b). Similar result is also obtained for WM which is shown in the SM~\cite{Note1}.

Going beyond the inversion symmetric materials discussed in this letter, the OMM-induced Hall effect is also present in materials which intrinsically break TRS. However, in these systems the Hall current has an additional intrinsic $B$- independent anomalous Hall and extrinsic $B$-dependent contributions arising from the BC.  We derive the expressions for all these contributions explicitly in the SM~\cite{Note1}. One simplification used in our work is a constant $\tau$ approximation. However, the energy or momentum dependence of $\tau$ can be easily incorporated in our formulation without changing the results qualitatively. 

To summarize, we have predicted novel OMM-induced {\it intrinsic} $B$-linear Hall effect in all the conductivities. These  are purely quantum mechanical in nature and they have no classical analog. These OMM-induced Hall effects originate from the geometric properties of the electronic wave-functions (the BC and the OMM) and are the only dissipation-less Hall effect in a non-centrosymmetric  material. 
In contrast to the Lorentz force's contributions, the OMM-induced Hall conductivities dominate near the band edges or band crossings, which are `hotspots' of the BC and the OMM, and these decrease on increasing the carrier concentration. These OMM-induced Hall components significantly impact the transverse MR measurements, giving rise to a negative MR$_\rho$ (quadratic in $B$) and an increase in the strength of positive MR in the Seebeck effect. 
We also show that the OMM-induced Hall conductivity leads to a significant under-estimation in the charge carrier density measured in the Hall resistivity experiments. 

\bibliography{OMM_HE_v7.bib}

\end{document}